\documentstyle[12pt]{article}
\begin{document}
\begin{center}
{\Large \bf
Indeterministic Quantum Gravity} \\[0.5cm]
{\large\bf III. Gravidynamics versus Geometrodynamics:
Revision of the Einstein Equation} \\[1.5cm]
{\bf Vladimir S.~MASHKEVICH}\footnote {E-mail:
mashkevich@gluk.apc.org}  \\[1.4cm]
{\it Institute of Physics, National academy
of sciences of Ukraine \\
252028 Kiev, Ukraine} \\[1.4cm]
\vskip 1cm

{\large \bf Abstract}
\end{center}
This paper is a continuation of the papers \cite{1,2}. A revision
of the Einstein equation shows that its dynamic incompleteness,
contrary to a popular opinion, cannot be circumvented by so-called
coordinate conditions. Gravidynamics, i.e., dynamics for
gravitational potentials $g_{\mu\nu}$ is advanced, which differs
from geometrodynamics of general relativity in that the former
is based on a projected Einstein equation. Cosmic gravidynamics,
due to a global structure of spacetime, is complete. The most
important result is a possibility of the closed universe with
a density below the critical one. \\[2cm]
Keywords: general relativity, cosmic time, cosmic space,
Einstein equation, quantum, metric, indeterministic

\newpage

\hspace*{6 cm}
\begin{minipage} [b] {10 cm}
\begin{flushright}
And have they fixed the where and when? \\
R.S. Hawker\vspace*{0.8 cm}
\end{flushright}
\end{minipage}

\begin{flushleft}
\hspace*{0.5 cm} {\Large \bf Introduction}
\end{flushleft}

As is generally known, the Einstein equations $G^{\mu\nu}=
T^{\mu\nu}$ have a peculiar feature from the standpoint of the
Cauchy problem: It is only the equations $G^{ij}=T^{ij}$ that
are dynamic ones, whereas the equations $G^{0\mu}=T^{0\mu}$
should be considered as constraints. The incompleteness of
the dynamic equations is circumvented by introducing coordinate
conditions.

On the other hand, in indeterministic quantum gravity the
equations $G^{ij}=T^{ij}$ are dynamically complete, whereas
the equations $G^{0\mu}=T^{0\mu}$ are absent at all. This is
due to quantum jumps and a global structure of spacetime
connected with them.

Thus, there is an essential divergence of spacetime dynamics
in general relativity and indeterministic quantum gravity.
The main objective of this paper is to elucidate the cause of
the divergence and to advance general spacetime dynamics.

First and foremost, we evidence that the circumvention of the
dynamic incompleteness of the Einstein equation by coordinate
conditions is a fallacy. The reasoning behind these conditions
is as follows. It is possible to pick an arbitrary coordinate
system, which involves four arbitrary functions. This
compensates for the lack of four dynamic equations. But this
construction, giving values for metric components $g_{\mu\nu}$,
does not determine to which events, i.e., spacetime points
these values correspond. Thus, spacetime dynamics of general
relativity, or so-called geometrodynamics \cite{3}, which is based
on the Einstein equation solely, is dynamically incomplete.

A complete spacetime dynamics should be based on the projected
(on a spacelike hypersurface) Einstein equation $G^{ij}=T^{ij}$
and some additional geometric conditions. Such a dynamics is
called gravidynamics---dynamics for gravitational potentials
$g_{\mu\nu}$.

In local problems, the complete covariant Einstein equation
$G^{\mu\nu}=T^{\mu\nu}$ must be fulfilled; so that in local
gravidynamics the equations $G^{0\mu}=T^{0\mu}$ should result
from the basic ones $G^{ij}=T^{ij}$.

In  global problems, the cosmic structure of spacetime suffices
for the basic equations to provide complete gravidynamics---cosmic
gravidynamics.

As local problems, the Schwarzschild solution and gravitational
collapse of a dust ball are considered. For these, the results
of general relativity and local gravidynamics coincide.

As a global problem, the Robertson-Walker spacetime and the
Friedmann universe are examined. The results of general
relativity and cosmodynamics, i.e., dynamics of the universe
based on cosmic gravidynamics, differ. The most important
conclusion of cosmodynamics is a possibility of the closed
universe with a density below the critical one.

\section{Unidynamics: Unified dynamics of spacetime and matter}
\subsection{Spacetime and matter}

We assume the universe as a physical system to be a pair
$(st,m)$, where $st$ is spacetime, $m$ is matter. In keeping
with general relativity as close as possible, we describe
spacetime classically and set
$$
st=(M,g,\nabla),\eqno (1.1)
$$
where $M$ is a differentiable 4-dimensional manifold, $g$ is
a metric, and $\nabla$ is the Levi-Civita  connection.

The matter may be assumed to be a family of fields (classical
or quantum) on $M$.

\subsection{Spacetime and matter dynamics}

Dynamics for the universe or its part is a unified dynamics of
spacetime and matter, which will be called unidynamics. Since
spacetime is described classically, matter dynamics is that of
matter fields on a given spacetime. Because matter dynamics
depends substantially on the description of the matter, our
main concern will be with spacetime dynamics. Dynamic variables
of spacetime are represented by a metric, so that spacetime
dynamics is that of the metric for a given time evolution of
the matter. In the sense of the aforesaid, spacetime dynamics and
that of matter are not entangled with each other.

\subsection{Local and global structures and problems}

A spacetime manifold $M$ possesses both local and global
structures. The local structure in a neighborhood of a point
is determined by the values of the metric $g$ in the
neighborhood. An important element of the global structure is
a spacelike hypersurface.

It is also reasonable to distinguish between local and global problems
of spacetime and unified dynamics. An example of the local
problem is the Schwarzschild solution (albeit it is formally
defined on an infinite domain). There is, in fact, the unique
global problem---that of the universe and its dynamics.

A solution to a local problem may involve only a local structure
of spacetime; a solution to the global problem will
involve a global structure as well.

\subsection{The principle of covariance and the geometric principle}

In the local structures and problems, an essential role is
played by the principle of covariance: Equations should be
phrased in a covariant tensor form. To include global elements,
it is efficient to advance the geometric principle: Spacetime
structure and dynamic equations should be phrased in geometric
form. The principle of covariance is a local form of the
geometric principle.

\section{Revision of the Einstein equation}
\subsection{Geometrodynamics: Spacetime dynamics in general relativity}

In general relativity, spacetime dynamics, which may be called
geometrody\-namics \cite{3}, is based on the Einstein equation
$$
G=T, \eqno (2.1)
$$
or, in components,
$$
G^{\mu\nu}=T^{\mu\nu},\quad \mu,\nu=0,1,2,3, \eqno (2.2)
$$
where $G$ is the Einstein tensor, $T$ is the energy-momentum
tensor. The Einstein equation is local, so geometrodynamics
is also local.

\subsection{The structure of the Einstein equation}

It is common knowledge that the system of eqs.(2.2) has peculiar
features from the standpoint of dynamics, i.e., time evolution:
The equations
$$
G^{0\mu}=T^{0\mu},\quad \mu=0,1,2,3, \eqno(2.3)
$$
do not contain the second time derivatives $\partial^2g_{\mu\nu}
/\partial t^2\;(t=x^0)$; the equations
$$
G^{ij}=T^{ij},\quad i,j=1,2,3, \eqno (2.4)
$$
contain the second time derivatives only for the $g_{ij}$. It
follows that there are only six dynamic equations (2.4) for
10 components $g_{\mu\nu}$, whereas the equations (2.3) are
constraints on the initial data. More specifically, eqs.(2.4)
determine the six derivatives $\partial^2g_{ij}/\partial t^2$,
but leave the remaining four, $\partial^2g_{0\mu}/\partial t^2$,
undetermined.

\subsection{Involvement of matter dynamics in geometrodynamics}

The constraints (2.3) may be used as a part of equations of
matter dynamics, which implies an involvement of the latter
in geometrodynamics, or an entangle\-ment of spacetime dynamics
and matter dynamics with each other. This entangle\-ment may be
expressed in terms of the equation
$$
T^{\mu\nu}{}_{;\nu}=0, \eqno(2.5)
$$
which follows from eq.(2.2). The entanglement imposes
restrictions on matter dynamics.

\subsection{Coordinate conditions}

The universally accepted way of circumventing the incompleteness
of the dynamic equations (2.4) consists in introducing
coordinate conditions (see, e.g., \cite{4}). The reasoning behind
these conditions is as follows. Let $x$ and $\bar x$ be two
coordinate systems:
$$
M\supset N\ni p\leftrightarrow x\leftrightarrow\bar x,\eqno(2.6)
$$
where $p$ is a point of spacetime, i.e., an event. We have
$$
\bar g^{0\mu}(\bar x)=\frac{\partial\bar x^0}{\partial x^\rho}
\frac{\partial\bar x^\mu}{\partial x^\sigma}g^{\rho\sigma}(x).
\eqno(2.7)
$$
There are four arbitrary functions
$$
\bar x^\mu=\bar x^\mu(x), \eqno(2.8)
$$
which makes it possible to prescribe given values for
$\bar g^{0\mu}$. This compensates for the lack of four
dynamic equations.

\subsection{Incompleteness of geometrodynamics}

We affirm that the reasoning outlined above is a fallacy. The
problem is to determine $g_p$---metric as a function of
spacetime point, or event $p$. (This function may be measured
\cite{5}.) It is important that the event is independent of the
metric. To emphasize this, we quote \cite{3}: ``...nature provides
its own way to localize a point in spacetime, as Einstein was
the first to emphasize. Characterize the point by what happens
there! Give a point in spacetime the name 'event'... The
primitive concept of an event... needs no refinement. The
essential property here is identifiability, which is not
dependent on the Lorentz metric structure of spacetime.''

Let some values for $\bar g^{0\mu}(\bar x)$ be prescribed.
Then the problem is to find the functions $\bar x^\mu(p)$. Let some
functions $x^\rho(p)$ be given. Then the problem reduces to
finding the functions $\bar x^\mu(x^\rho)$. But the functions
$g^{\rho\sigma}(x)$ are not known, so we cannot find
$\bar x^\mu(x^\rho)$ from (2.7).

Here is a parody of the situation outlined above. Let $v(x)$
be a contravariant vector field on the real axis $-\infty<
x<\infty$. Let $\bar x(x)$ be a new coordinate, then
$$
\bar v(\bar x)=\frac{d\bar x}{dx}v(x). \eqno(2.9)
$$
For a given $v(x)$, a prescription for $\bar v(\bar x)$
determines a function $\bar x(x)$: (2.9) is an equation for
the latter. But if $v(x)$ is not known, $\bar x(x)$
is also unknown, so $\bar v(\bar x)$ gives in fact no
information on $v(x)$.

Thus, so-called coordinate conditions do not circumvent the
incompleteness of geometro\-dynamics.

\section{Gravidynamics}
\subsection{Projection of the Einstein equation}

In constructing a complete spacetime dynamics, we first of all
disentangle spacetime and matter dynamics from each other.
Namely, in the general case, we discard the constraints (2.3).
The grounds for this are as follows. In a quantum description of
matter, eqs.(2.3) take the form
$$
G^{0\mu}=(\Psi,T^{0\mu}\Psi), \eqno(3.1)
$$
where $T$ is the operator of the energy-momentum tensor, $\Psi$
is a state vector for matter. The components $G^{0\mu}$ do not
involve the second time derivatives $\partial^2g^{\mu\nu}/
\partial t^2$, so that in indeterministic dynamics eqs.(3.1)
are violated at quantum jumps, i.e., jumps of $\Psi$ \cite{2}.

Thus, generally, we retain only the dynamic equations (2.4). These
represent a projection of the complete Einstein equation (2.2)
on a spacelike hypersurface.

\subsection{Gravidynamics: Unconstrained geometrodynamics}

As spacetime dynamics we will use unconstrained
geometrodynamics, i.e., dynamics based on the projected
Einstein equation (2.4). That is dynamics for gravitational
potentials $g_{\mu\nu}$, in view of which we call it
gravidynamics. Thus gravidynamics is spacetime dynamics based
on the projected Einstein equation.

\subsection{Problems of completeness and geometric formulation}

The projected Einstein equation (2.4) neither provides
completeness of spacetime dynamics nor imparts a geometric
formulation to it. So a proposal for using gravidynamics as
a spacetime dynamics gives rise immediately to problems
concerned with completeness and the geometric principle.
These problems should be dealt with separately for local and
global dynamics.

\subsection{Local gravidynamics}

Local gravidynamics should generally involve no global
structure. This implies that for it the geometric principle
boils down to the local form, i.e., the principle of covariance.
But a covariant equation corresponding to the projected Einstein
equation (2.4) is the complete Einstein equation (2.2). It
follows that in local gravidynamics the constraints (2.3) should
hold, but they must be a consequence of the dynamic equations
(2.4).

As for completeness of local gravidynamics, it should be provided
in every concrete case by additional physical and/or geometrical
conditions.

\subsection{Global structure: Cosmic time and space}

Global gravidynamics, in addition to the projected Einstein
equation (2.4), should involve a global structure of the
spacetime manifold $M$, so that a complete spacetime dynamics
would have been obtained. The global structure is suggested
by, firstly, the Robertson-Walker spacetime and, secondly,
indetermin\-istic quantum gravity, which incorporates quantum
jumps giving rise to cosmic time. The structure is as follows.
The manifold M is a direct product of two manifolds:
$$
M=T\times S,\qquad M\ni p=(t,s),\quad t\in T,\quad s\in S.
\eqno(3.2)
$$
The 1-dimensional manifold $T$ is the cosmic time, the
3-dimensional one $S$ is the cosmic space. By eq.(3.2) the
tangent space at a point $p\in M$ is
$$
M_p=T_p\oplus S_p. \eqno(3.3)
$$
Next it is assumed that
$$
T\perp S. \eqno(3.4)
$$
It follows for the metric tensor
$$
g=g_T+g_S. \eqno(3.5)
$$
Furthermore,
$$
g=dt\otimes dt-\tilde g_t, \eqno(3.6)
$$
which gives a geometric, i.e., coordinateless representation
for $g$. A relevant coordinate representation is
$$
g=dt^2-\tilde g_{t\,ij}dx^idx^j,\quad g_{0\mu}=\delta_{0\mu},
\quad g_{ij}=-\tilde g_{t\,ij}. \eqno(3.7)
$$

\subsection{Cosmic gravidynamics}

The equation of motion for the metric is the projected Einstein
equation
$$
G_S=T_S, \eqno(3.8)
$$
or eq.(2.4). There are six equations for six quantities
$\tilde g_{ij}$, so that spacetime dynamics is complete.
It is natural to call this dynamics cosmic gravidynamics.

Note that the form (3.7) for the metric holds in any Gaussian
coordinates, but in cosmic gravidynamics the projected
Einstein equation (2.4), or (3.8), is generally fulfilled
only in cosmic representation given by eqs.(3.2)-(3.5).

\section{On the problem of matter dynamics}

In this paper, the main line of investigation concerns
spacetime dynamics. As for matter dynamics, we restrict the
consideration to a selection of problems.

In local gravidynamics, the complete Einstein equation (2.2)
holds. There\-fore, local matter dynamics should satisfy the
equation (2.5) of a vanishing covariant divergence.

Global matter dynamics may or may not satisfy eq.(2.5). This
equation does not imply the complete Einstein equation (2.2):
The constraints (2.3) are not generally fulfilled.

\section{Cosmodynamics}
\subsection{Cosmic unidynamics}

Dynamics for the universe consists of a global spacetime
dynamics and a matter dynamics. As the global spacetime dynamics
we assume cosmic gravidynamics, so that the corresponding
unidynamics may be called cosmic one, or, for the sake of
brevity, cosmodynamics. The problem of constructing the
latter is that of matter dynamics.

\subsection{Contraposition with general relativity}

General relativity per se is a local theory, which involves generally
no additional global structure of the spacetime manifold.
Spacetime dynamics (geometrodynamics) is based on the complete
Einstein equation (2.1). Spacetime and matter dynamics are
entangled with each other. Geometrodynamics is incomplete.

Cosmodynamics is a global theory, which involves an additional
global structure of the spacetime manifold given by
eqs.(3.2)-(3.4). Spacetime dynamics (gravidynamics) is based
on the projected Einstein equation (3.8). Spacetime and matter
dynamics are disentangled from each other. Cosmic gravidynamics
is complete.

\subsection{Indeterministic cosmodynamics: Indeterministic quantum
grav\-ity as cosmodynamics}

Indeterministic quantum gravity, which is being developed in
this series of papers, is based on cosmic gravidynamics. Matter
dynamics is an indeterministic quantum one. So indeterministic
quantum gravity may be called indeterministic cosmodynamics
(indeterministic implies quantum).

\section{The Schwarzschild solution and gravitational collapse of a
dust ball}
\subsection{Treatment in general relativity}

By way of example of a local problem, we consider the
Schwarzschild solution and gravitational collapse of a dust
ball. Our initial concern will be the treatment in general
relativity (see, e.g., \cite{6,4}), or in geometrodynamics.

In the Schwarzschild solution the metric has the form
$$
g=e^{\nu(r)}dt^2-e^{\lambda(r)}dr^2-r^2(d\theta^2+\sin^2
\!\theta\,d\varphi^2). \eqno(6.1)
$$
Nontrivial components of the complete Einstein equation
are
$$
G^1_1=0\Rightarrow-e^{-\lambda}\left(\frac{\nu'}{r}+\frac
{1}{r^2}\right)+\frac{1}{r^2}=0, \eqno(6.2)
$$
$$
G^2_2=G^3_3=0\Rightarrow\nu''+\frac{\nu'^2}{2}+
\frac{\nu'-\lambda'}{r}-\frac{\nu'\lambda'}{2}=0,
\eqno(6.3)
$$
$$
G^0_0=0\Rightarrow-e^{-\lambda}\left(\frac{1}{r^2}-
\frac{\lambda'}{r}\right)+\frac{1}{r^2}=0, \eqno(6.4)
$$
prime denotes $d/dr$.

For the dust ball, the metric is
$$
g=dt^2-R^2(t)\left[\frac{dr^2}{1-kr^2}+r^2d\theta^2+
r^2\sin^2\!\theta\,d\varphi^2\right]. \eqno(6.5)
$$
The complete Einstein equation gives
$$
G_{ij}=8\pi\kappa T_{ij}\Rightarrow k+2R\ddot R+
\dot R^2=0, \eqno(6.6)
$$
$$
G_{0\mu}=8\pi\kappa T_{0\mu}\Rightarrow R(\dot R^2+k)=
k, \eqno(6.7)
$$
where the gravitational constant $\kappa$ is introduced,
point denotes $d/dt$,
$$
R(0)=1,\quad \dot R(0)=0, \eqno(6.8)
$$
$$
k=\frac{8\pi\kappa}{3}\rho(0), \eqno(6.9)
$$
$$
\rho(t)=\rho(0)/R^3(t) \eqno(6.10)
$$
is the matter density.

Gluing together the interior and exterior solutions leads
to
$$
k=\frac{2M\kappa}{a^3}, \eqno(6.11)
$$
where $a$ is the ball radius in the comoving coordinate
system,
$$
M=\frac{4\pi}{3}\rho(0)a^3 \eqno(6.12)
$$
is the ball mass.

\subsection{Treatment in local gravidynamics}

In local gravidynamics, for the Schwarzschild problem
eqs.(6.2),(6.3) hold, but eq.(6.4) is not given. Differentiating
eq.(6.2) we obtain
$$
\nu''+\frac{\nu'-\lambda'}{r}=\nu'\lambda',\eqno(6.13)
$$
from which and eq.(6.3) it follows
$$
\nu'(\nu'+\lambda')=0.\eqno(6.14)
$$

The solution
$$
\nu'=0,\qquad\nu={\rm const},\eqno(6.15)
$$
leads, by eq.(6.2), to
$$
\lambda=0,\eqno(6.16)
$$
so that the metric (6.1) is reduced, by $e^\nu dt^2\to dt^2$,
to the Minkowski one.

The solution
$$
\nu'+\lambda'=0\eqno(6.17)
$$
to the eq.(6.14) reduces eq.(6.2) to eq.(6.4). Thus the
constraint (6.4) is a consequence of the projected Einstein
equation (6.2),(6.3), so that the Schwarz\-schild solution
holds in local gravidynamics.

For the dust ball, eq.(6.6) is valid, but eq.(6.7) and the
value (6.9) for $k$ are not given. We obtain from eq.(6.6)
$$
\frac{d}{dR}[R(k+\dot R^2)]=0,\qquad R(k+\dot R^2)=C.
\eqno(6.18)
$$
{}From the initial conditions (6.8) it follows $C=k$, so
that eq.(6.18) results in eq.(6.7). Gluing together the
interior and exterior solutions leads again to the value
(6.11), which, combined with eq.(6.12), reduces to the
expression (6.9). Thus, local gravidynamics reproduces
the results of geometrodynamics.

\subsection{Agreement of the results}

For the local problems considered, the results given by
gravidynamics and geometrodynamics coincide. The essence of
the matter is that for these problems, the constraints are
consequences of the projected Einstein equation.

\section{The Robertson-Walker spacetime and the Friedmann universe}
\subsection{Treatment in general relativity}

As a global problem, we consider the Robertson-Walker
spacetime and the Friedmann universe. We begin with the
treatment in general relativity (see, e.g., \cite{4}).

The Robertson-Walker metric is of the form
$$
g=dt^2-R^2(t)\left\{\frac{dr^2}{1-kr^2}+r^2d\theta^2+r^2
\sin^2\theta\,d\varphi^2\right\},\quad k=-1,0,1.\eqno(7.1)
$$
For the Friedmann universe, the complete Einstein equation
gives
$$
G_{ij}=8\pi\kappa T_{ij}\Rightarrow 2\ddot RR+\dot R^2+k=
-8\pi\kappa pR^2,\eqno(7.2)
$$
$$
G_{0\mu}=8\pi\kappa T_{0\mu}\Rightarrow \dot R^2+k=
\frac{8\pi\kappa}{3}\rho R^2,\eqno(7.3)
$$
where $p$ is the pressure.

{}From eqs.(7.2),(7.3) it follows
$$
\dot\rho R^3+3(\rho+p)R^2\dot R=0,\eqno(7.4)
$$
which is equivalent to
$$
T^{0\nu}{}_{;\nu}=0,\eqno(7.5)
$$
or
$$
dE=-pdV,\quad V\sim R^3.\eqno(7.6)
$$

\subsection{Treatment in cosmodynamics}

In cosmodynamics, the dynamic equation (7.2) holds, but the
constraint (7.3) is not given. Let an equation for matter be
(7.5), or (7.6). We obtain from (7.6)
$$
pR^2=-\frac{1}{3}\frac{d(\rho R^3)}{dR}.\eqno(7.7)
$$
This equation leads to eq.(7.4).

We obtain from (7.2)
$$
\frac{d}{dR}(R\dot R^2+kR)=-8\pi\kappa pR^2,\eqno(7.8)
$$
which, combined with eq.(7.7), results in
$$
\frac{d}{dR}\left(R\dot R^2+kR-\frac{8\pi\kappa}{3}\rho R^3
\right)=0,\eqno(7.9)
$$
i.e.,
$$
R\dot R^2+kR-\frac{8\pi\kappa}{3}\rho R^3=L={\rm const}.\eqno(7.10)
$$

\subsection{Contraposition}

In cosmodynamics, there exists an integral of motion $L$ given
by eq.(7.10). In general relativity, the only possible value
for this integral is, by eq.(7.3), zero:
$$
L=0.\eqno(7.11)
$$
Thus, in cosmodynamics the constraint (7.3) does not hold. The
condition $L={\rm const}$ is substantially less restrictive than
$L=0$. So cosmodynamics gives a much more general result for the
Friedmann model than the result of general relativity.

\subsection{On critical density}

The constraint (7.3) may be written in the form
$$
\frac{\rho-\rho_c}{\rho_c}=\frac{k}{\dot R^2}\,,\eqno(7.12)
$$
where
$$
\rho_c=\frac {3}{8\pi\kappa}\left(\frac{\dot R}{R}\right)^2=
\frac{3}{8\pi\kappa}H^2 \eqno(7.13)
$$
is the critical density and $H$ is Hubble's constant.

In cosmodynamics, we obtain from eq.(7.10), in place of
(7.12),
$$
\frac{\rho-\rho_c}{\rho_c}=\frac{kR-L}{R\dot R^2}.
\eqno(7.14)
$$
This important result implies, specifically, that there is a
possibility of a closed universe $(k=1)$ with a density
$\rho$ which is smaller than the critical one, $\rho_c$. This
is realized as long as the inequality
$$
R<L \eqno(7.15)
$$
is fulfilled.

\end{document}